\documentclass[showpacs,aps,prd]{revtex4}
\input epsf

\begin{document}

\title{Improved QCD sum rule study of $Z_{c}(3900)$ as a $\bar{D}D^{*}$ molecular state}
\author{Jian-Rong Zhang}
\affiliation{Department of Physics, College of Science, National University of Defense Technology,
Changsha 410073, Hunan, People's Republic of China}

%\date{\today}

%%%%%%%%%%%%%%%%%%%%%%%%%%%%%%%%%%%%%%%%%%%%%%%%%%%%%%%%%%%%%%%%%%%%%
\begin{abstract}
In the framework of QCD sum rules, we present an improved study of our previous work [Phys. Rev. D {\bf80}, 056004 (2009)]
particularly on the $\bar{D}D^{*}$ molecular state to
investigate that the possibility of the newly observed $Z_{c}(3900)$ as a $S$-wave $\bar{D}D^{*}$ molecular state.
To ensure the quality of QCD sum rule analysis, contributions of
up to dimension nine are calculated to test the convergence of  operator product expansion (OPE).
We find that the two-quark condensate $\langle\bar{q}q\rangle$
is very large and makes the standard OPE convergence (i.e. the perturbative at least
larger than each condensate contribution) happen at very large values of Borel parameters.
By releasing the rigid OPE convergence criterion, one could find that the OPE convergence is still under control.
We arrive at
the numerical result $3.86\pm0.27~\mbox{GeV}$ for
$\bar{D}D^{*}$,
which agrees with the mass of $Z_{c}(3900)$ and could support
the explanation of $Z_{c}(3900)$
in terms of a $S$-wave $\bar{D}D^{*}$ molecular state.
\end{abstract}
\pacs {11.55.Hx, 12.38.Lg, 12.39.Mk}\maketitle

\section{Introduction}\label{sec1}
Very recently, BESIII Collaboration studied the process $e^{+}e^{-}\rightarrow\pi^{+}\pi^{-}J/\psi$ at a center-of-mass energy of $4.26~\mbox{GeV}$, and reported the observation of a new charged charmonium-like
structure $Z_{c}(3900)$ in the $\pi^{\pm}J/\psi$ invariant spectrum with a mass of $3899.0\pm3.6\pm4.9~\mbox{MeV}$ and a width
of $46\pm10\pm20~\mbox{MeV}$ \cite{3900-BES}.
Before the BESIII's observation, Chen {\it et al.} predicted
that a charged
charmonium-like structure is observable
in the $Y(4260)\rightarrow J/\psi\pi^{+}\pi^{-}$ process \cite{Chen}.
In the study of $Y(4260)\rightarrow\pi^{+}\pi^{-}J/\psi$ decays,
Belle Collaboration also observed
a $Z(3895)^{\pm}$ state with a mass of $3894.5\pm6.6\pm4.5~\mbox{MeV}$ and a width of
$63\pm24\pm26~\mbox{MeV}$ in the $\pi^{\pm}J/\psi$ mass spectrum \cite{3900-Belle}.
Xiao {\it et al.} confirmed the charged state $Z_{c}(3900)$
in the analysis of data taken with the CLEO-c detector at $\psi(4160)$, and measured its mass and width to be
$3885\pm5\pm1~\mbox{MeV}$ and $34\pm12\pm4~\mbox{MeV}$, respectively \cite{3900-CLEO}.

The new experimental results have aroused theorists' great
interest in comprehending the internal structures of $Z_{c}(3900)$.
Soon after the observation of $Z_{c}(3900)$, it was proposed that these states are $S$-wave $\bar{D}D^{*}$ molecules \cite{Zhao,Guo}.
Subsequently, there also appeared many other works to explain this exotic states \cite{Chen1,Maiani,Karliner,Voloshin,Mahajan,Wilbring,Li}.
Undoubtedly, it is interesting and significative to investigate that whether $Z_{c}(3900)$
could be a $S$-wave $\bar{D}D^{*}$ state.
To understand the inner structure of $Z_{c}(3900)$,
it is very helpful and quite needed to determine their properties like masses quantitatively.
Nowadays, QCD is widely believed to be the true theory of describing strong interactions.
However, it is quite difficult to acquire the hadron spectrum from QCD first principles.
The main reason is that low energy QCD involves a regime where it is futile to attempt perturbative calculations
and the strong interaction dynamics of hadronic systems
is governed by nonperturbative QCD effects completely.
Meanwhile, one has limited knowledge on nonperturbative QCD aspects
for there are still many questions remain
unanswered or realized only at a qualitative level.

The method of
QCD sum rules \cite{svzsum} is a nonperturbative
formulation firmly based on the basic theory of QCD,
which
has been successfully applied
to conventional hadrons (for reviews
see \cite{overview1,overview2,overview3,overview4} and references
therein) and multiquark states (e.g.
see \cite{XYZ}).
In particular for the $S$-wave $\bar{D}D^{*}$ molecular state,
we have definitely predicted its mass to be $3.88\pm0.10~\mbox{GeV}$ with QCD sum rules
several years ago in Ref. \cite{Previous}, in which mass spectra of molecular states with various $\{Q\bar{q}\}\{\bar{Q}^{(')}q\}$ configurations
have been systematically studied.
Numerically, one could see
that our prediction for the mass of $\bar{D}D^{*}$ state agrees well
with the experimental data of newly observed $Z_{c}(3900)$.
That result could support the explanation of $Z_{c}(3900)$ as a $S$-wave $\bar{D}D^{*}$ molecular state.
At present, we would put forward an improved study of our previous work
on the $\bar{D}D^{*}$ state in view of below reasons.
First, it is known that one can analyze the OPE convergence and
the pole contribution dominance to determine the conventional
Borel window in the standard QCD sum rule
approach to ensure the validity of QCD sum rule analysis.
However, we find that it may be
difficult to find a conventional work window rigidly satisfying
both of the two rules in some recent works \cite{Zs}, which actually has also been discussed in
some other works (e.g., Refs. \cite{Matheus,HXChen,ZGWang}).
The main reason is that some high dimension condensates
are very large and play an important role in the OPE
side, which makes the standard OPE convergence happen only at very large values of Borel parameters.
Whereas in the previous Ref. \cite{Previous},
we merely considered
contributions of the operators up to dimension six in OPE
and the Borel windows are roughly taken the same values for the
similar class of states for simplicity and convenience.
Thus, it may be more reliable to test the OPE convergence by including
higher dimension condensate contributions than six and considering
the work windows minutely, and then one could more
safely extract the hadronic information from QCD sum rules.
Second, even higher condensate contributions
may not radically influence the character of OPE convergence in some case,
one still could attempt to improve the theoretical result because
some higher condensates are helpful to stabilize the Borel curves.
Particularly for the newly observed $Z_{c}(3900)$ states, they can not be simple $c\bar{c}$ conventional mesons
since they are electric charged. It may be a new hint for the existence of exotic hadrons
and $Z_{c}(3900)$ are some ideal candidates for them.
Once exotic states can be confirmed by experiment, QCD will be further tested
and then one will understand QCD low-energy behaviors more
deeply.
Therefore,
it is of importance and worth to make meticulous theoretical
efforts to reveal the underlying structures of $Z_{c}(3900)$. All in all, we would like to improve our previous work
to investigate that whether $Z_{c}(3900)$ could serve as a $\bar{D}D^{*}$ molecular state.

The rest of the paper is organized as follows. In Sec. \ref{sec2}, QCD sum
rules for the molecular states are introduced, and both the
phenomenological representation and QCD side are derived, followed
by the numerical analysis and some discussions in Sec. \ref{sec3}.
The last Section is a brief
summary.
%%%%%%%%%%%%%%%%%%%%%%%%%%%%%%%%%%%%%%%%%%%%%%%%%%%%%%%%%%%%%%%%%%%
\section{molecular state QCD sum rules}\label{sec2}
The starting point of the QCD sum rule method is to construct a proper
interpolating current to represent the studied state.
One knows that the method of
QCD sum rules
has been widely applied
to multiquark systems since the experimental observations of many
new hadrons in these years.
At present, currents of molecular states and tetraquark states
could be differentiated by their different construction ways.
Concretely, molecular currents are built up with the color-singlet currents of
their composed hadrons to form hadron-hadron
configurations of fields, which are different
from currents of tetraquark states constructed by diquark-antidiquark configurations
of fields. What's needed to note
is that these two types of currents can be related to
each other by Fiertz rearrangement.
However, the transformation relations are suppressed by corresponding color and Dirac
factors \cite{XYZ} and one
could obtain a reliable sum rule while choosing
the appropriate current to represent
the physical state.
This means that if the
physical state is a molecular state, it would be best to choose a meson-meson type of current so that it has a large overlap with the
physical state. Similarly for a tetraquark state, it would be best to choose a diquark-antidiquark type of current. When the
sum rule reproduces a mass consistent well with the physical value, one can infer that the physical state has a structure well
represented by the chosen current. In this way, one can indirectly and commonly discriminate between the molecular and the tetraquark
structures of observed states.
One can expect that the judgements could be very effective for some ideal cases,
e.g. the results obtained from different types of currents are very different
so that one could easily discriminate them.
Note that in some exceptional cases that it may not be very different
for final results from molecular currents and tetraquark currents.
For example, Narison {\it et al.} investigated both molecular and tetraquark currents associated with
$X(3872)$ and they finally gained the same mass predictions within the
accuracy of QCD sum rule method in Ref. \cite{3872}.
For the present work, in order to
study the possibility of $Z_{c}(3900)$ as a $S$-wave $\bar{D}D^{*}$ molecular state,
we thus construct the molecular current from corresponding
 currents of $\bar{D}$ and $D^{*}$ mesons to form meson-meson
configurations of fields.
In full theory, the interpolating currents for heavy $D$
mesons can be found in Ref. \cite{reinders}.
Therefore, one can build the following form of current
\begin{eqnarray}
j^{\mu}_{\bar{D}D^{*}}&=&(\bar{Q}_{a}i\gamma_{5}q_{a})(\bar{q}_{b}\gamma^{\mu}Q_{b}),\nonumber
\end{eqnarray}
for $\bar{D}D^{*}$ with $J^{P}=1^{+}$,
where $q$ indicates the light $u$ or $d$ quark, $Q$ denotes the heavy $c$ quark, and the subscript $a$
and $b$ are color indices.
Note that the quantum numbers of $Z_{c}(3900)$ have not been given
experimentally for the moment, and $1^{+}$ is just one possible choice of their spin-parities.

One can then write down the two-point correlator
\begin{eqnarray}
\Pi^{\mu\nu}(q^{2})=i\int
d^{4}x\mbox{e}^{iq.x}\langle0|T[j^{\mu}_{\bar{D}D^{*}}(x)j^{\nu+}_{\bar{D}D^{*}}(0)]|0\rangle.
\end{eqnarray}
Lorentz covariance implies that the correlator can be generally parameterized as
\begin{eqnarray}
\Pi^{\mu\nu}(q^{2})=\bigg(\frac{q^{\mu}q^{\nu}}{q^{2}}-g^{\mu\nu}\bigg)\Pi^{(1)}(q^{2})+\frac{q^{\mu}q^{\nu}}{q^{2}}\Pi^{(0)}(q^{2}).
\end{eqnarray}
The term proportional to $g_{\mu\nu}$ will be chosen to extract the
mass sum rule. Phenomenologically,
$\Pi^{(1)}(q^{2})$ can be expressed as
\begin{eqnarray}\label{ph}
\Pi^{(1)}(q^{2})=\frac{[\lambda^{(1)}]^{2}}{M_{\bar{D}D^{*}}^{2}-q^{2}}+\frac{1}{\pi}\int_{s_{0}}
^{\infty}ds\frac{\mbox{Im}\Pi^{(1)\mbox{phen}}(s)}{s-q^{2}}+\mbox{subtractions},
\end{eqnarray}
where $M_{\bar{D}D^{*}}$ denotes the mass of the $\bar{D}D^{*}$ state,
$s_0$ is the continuum threshold parameter, and $\lambda^{(1)}$ gives
the coupling of the current to the hadron $\langle0|j^{\mu}_{\bar{D}D^{*}}|\bar{D}D^{*}\rangle=\lambda^{(1)}\epsilon^{\mu}$.
In the OPE side, $\Pi^{(1)}(q^{2})$ can be written as
\begin{eqnarray}\label{ope}
\Pi^{(1)}(q^{2})=\int_{4m_{Q}^{2}}^{\infty}ds\frac{\rho^{\mbox{OPE}}(s)}{s-q^{2}}+\Pi_{1}^{\mbox{cond}}(q^{2}),
\end{eqnarray}
where the spectral density is given by
$\rho^{\mbox{OPE}}(s)=\frac{1}{\pi}\mbox{Im}\Pi^{\mbox{(1)}}(s)$.
Technically, we work at leading order in $\alpha_{s}$ and consider condensates up
to dimension nine, employing the similar techniques as Refs. \cite{Tech,Zhang}. To keep the heavy-quark mass finite, one can use the
momentum-space expression for the heavy-quark propagator \cite{reinders}
\begin{eqnarray}
S_{Q}(p)&=&\frac{i}{\rlap/p-m_{Q}}
-\frac{i}{4}gt^{A}G^{A}_{\kappa\lambda}(0)\frac{1}{(p^{2}-m_{Q}^{2})^{2}}[\sigma_{\kappa\lambda}(\rlap/p+m_{Q})
+(\rlap/p+m_{Q})\sigma_{\kappa\lambda}]\nonumber\\&&{}
-\frac{i}{4}g^{2}t^{A}t^{B}G^{A}_{\alpha\beta}(0)G^{B}_{\mu\nu}(0)\frac{\rlap/p+m_{Q}}{(p^{2}-m_{Q}^{2})^{5}}[
\gamma^{\alpha}(\rlap/p+m_{Q})\gamma^{\beta}(\rlap/p+m_{Q})\gamma^{\mu}(\rlap/p+m_{Q})\gamma^{\nu}\\&&{}
+\gamma^{\alpha}(\rlap/p+m_{Q})\gamma^{\mu}(\rlap/p+m_{Q})\gamma^{\beta}(\rlap/p+m_{Q})\gamma^{\nu}+
\gamma^{\alpha}(\rlap/p+m_{Q})\gamma^{\mu}(\rlap/p+m_{Q})\gamma^{\nu}(\rlap/p+m_{Q})\gamma^{\beta}](\rlap/p+m_{Q})\nonumber\\&&{}
+\frac{i}{48}g^{3}f^{ABC}G^{A}_{\gamma\delta}G^{B}_{\delta\varepsilon}G^{C}_{\varepsilon\gamma}\frac{1}{(p^{2}-m_{Q}^{2})^{6}}(\rlap/p+m_{Q})
[\rlap/p(p^{2}-3m_{Q}^{2})+2m_{Q}(2p^{2}-m_{Q}^{2})](\rlap/p+m_{Q}).\nonumber
\end{eqnarray}
The light-quark part of the
correlator can be calculated in the coordinate space, with the light-quark
propagator
\begin{eqnarray}
S_{ab}(x)&=&\frac{i\delta_{ab}}{2\pi^{2}x^{4}}\rlap/x-\frac{m_{q}\delta_{ab}}{4\pi^{2}x^{2}}-\frac{i}{32\pi^{2}x^{2}}t^{A}_{ab}gG^{A}_{\mu\nu}(\rlap/x\sigma^{\mu\nu}
+\sigma^{\mu\nu}\rlap/x)-\frac{\delta_{ab}}{12}\langle\bar{q}q\rangle+\frac{i\delta_{ab}}{48}m_{q}\langle\bar{q}q\rangle\rlap/x\nonumber\\&&{}\hspace{-0.3cm}
-\frac{x^{2}\delta_{ab}}{3\cdot2^{6}}\langle g\bar{q}\sigma\cdot Gq\rangle
+\frac{ix^{2}\delta_{ab}}{2^{7}\cdot3^{2}}m_{q}\langle g\bar{q}\sigma\cdot Gq\rangle\rlap/x-\frac{x^{4}\delta_{ab}}{2^{10}\cdot3^{3}}\langle\bar{q}q\rangle\langle g^{2}G^{2}\rangle,
\end{eqnarray}
which is then
Fourier-transformed to the momentum space in $D$ dimension.
Since masses of light $u$ and $d$ quarks are three order magnitudes less than the one of heavy $c$ quark,
they are neglected here following the usual treatment.
The
resulting light-quark part is combined with the heavy-quark part
before it is dimensionally regularized at $D=4$.
After equating Eqs. (\ref{ph}) and (\ref{ope}), assuming quark-hadron duality, and
making a Borel transform, the sum rule can be written as
\begin{eqnarray}\label{sumrule1}
[\lambda^{(1)}]^{2}e^{-M_{\bar{D}D^{*}}^{2}/M^{2}}&=&\int_{4m_{Q}^{2}}^{s_{0}}ds\rho^{\mbox{OPE}}e^{-s/M^{2}}+\hat{B}\Pi_{1}^{\mbox{cond}},
\end{eqnarray}
with $M^2$ the Borel parameter.
Making
the derivative in terms of $M^2$ to the sum rule and then dividing
by itself, we have the mass of $\bar{D}D^{*}$ state
\begin{eqnarray}\label{sum rule 1}
M_{\bar{D}D^{*}}^{2}&=&\bigg\{\int_{4m_{Q}^{2}}^{s_{0}}ds\rho^{\mbox{OPE}}s
e^{-s/M^{2}}+\frac{d\hat{B}\Pi_{1}^{\mbox{cond}}}{d(-\frac{1}{M^{2}})}\bigg\}/
\bigg\{\int_{4m_{Q}^{2}}^{s_{0}}ds\rho^{\mbox{OPE}}e^{-s/M^{2}}+\hat{B}\Pi_{1}^{\mbox{cond}}\bigg\},
\end{eqnarray}
where
\begin{eqnarray}
\rho^{\mbox{OPE}}(s)=\rho^{\mbox{pert}}(s)+\rho^{\langle\bar{q}q\rangle}(s)+\rho^{\langle
g^{2}G^{2}\rangle}(s)+\rho^{\langle
g\bar{q}\sigma\cdot G q\rangle}(s)+\rho^{\langle\bar{q}q\rangle^{2}}(s)+\rho^{\langle
g^{3}G^{3}\rangle}(s)+\rho^{\langle\bar{q}q\rangle\langle
g^{2}G^{2}\rangle}(s),\nonumber
\end{eqnarray}
with $\rho^{\mbox{pert}}$, $\rho^{\langle\bar{q}q\rangle}$, $\rho^{\langle
g^{2}G^{2}\rangle}$, $\rho^{\langle
g\bar{q}\sigma\cdot G q\rangle}$,
$\rho^{\langle\bar{q}q\rangle^{2}}$, $\rho^{\langle g^{3}G^{3}\rangle}$,
and $\rho^{\langle\bar{q}q\rangle\langle
g^{2}G^{2}\rangle}$ are the
perturbative, two-quark condensate, two-gluon condensate, mixed
condensate, four-quark condensate, three-gluon condensate, and two-quark multiply two-gluon condensate
spectral densities, respectively. In fact, the
spectral densities up to dimension six
have been given in our previous work \cite{Previous}, which
are also enclosed here for the paper's completeness. Concretely, the
spectral densities are
\begin{eqnarray}
\rho^{\mbox{pert}}(s)&=&\frac{3}{2^{12}\pi^{6}}\int_{\alpha_{min}}^{\alpha_{max}}\frac{d\alpha}{\alpha^{3}}\int_{\beta_{min}}^{1-\alpha}\frac{d\beta}{\beta^{3}}(1-\alpha-\beta)(1+\alpha+\beta)r(m_{Q},s)^{4},\nonumber\\
\rho^{\langle\bar{q}q\rangle}(s)&=&-\frac{3\langle\bar{q}q\rangle}{2^{7}\pi^{4}}m_{Q}\int_{\alpha_{min}}^{\alpha_{max}}\frac{d\alpha}{\alpha^{2}}\int_{\beta_{min}}^{1-\alpha}\frac{d\beta}{\beta}(1+\alpha+\beta)r(m_{Q},s)^{2},\nonumber\\
\rho^{\langle g^{2}G^{2}\rangle}(s)&=&\frac{\langle
g^{2}G^{2}\rangle}{2^{11}\pi^{6}}m_{Q}^{2}\int_{\alpha_{min}}^{\alpha_{max}}\frac{d\alpha}{\alpha^{3}}\int_{\beta_{min}}^{1-\alpha}d\beta(1-\alpha-\beta)(1+\alpha+\beta)r(m_{Q},s),\nonumber\\
\rho^{\langle g\bar{q}\sigma\cdot G q\rangle}(s)&=&\frac{3\langle
g\bar{q}\sigma\cdot G
q\rangle}{2^{8}\pi^{4}}m_{Q}\int_{\alpha_{min}}^{\alpha_{max}}d\alpha\bigg\{\int_{\beta_{min}}^{1-\alpha}\frac{d\beta}{\beta}r(m_{Q},s)-\frac{2}{1-\alpha}[m_{Q}^{2}-\alpha(1-\alpha)
s]\bigg\},\nonumber\\
\rho^{\langle\bar{q}q\rangle^{2}}(s)&=&\frac{\langle\bar{q}q\rangle^{2}}{2^{4}\pi^{2}}m_{Q}^{2}\sqrt{1-\frac{4m_{Q}^{2}}{s}},\nonumber\\
\rho^{\langle g^{3}G^{3}\rangle}(s)&=&\frac{\langle
g^{3}G^{3}\rangle}{2^{13}\pi^{6}}\int_{\alpha_{min}}^{\alpha_{max}}\frac{d\alpha}{\alpha^{3}}\int_{\beta_{min}}^{1-\alpha}d\beta(1-\alpha-\beta)(1+\alpha+\beta)[r(m_{Q},s)+2
m_{Q}^{2}\beta],\nonumber\\
\rho^{\langle\bar{q}q\rangle\langle
g^{2}G^{2}\rangle}(s)&=&-\frac{\langle\bar{q}q\rangle\langle
g^{2}G^{2}\rangle}{2^{11}\pi^{4}}m_{Q}\bigg[\sqrt{1-\frac{4m_{Q}^{2}}{s}}+4\int_{\alpha_{min}}^{\alpha_{max}}\frac{d\alpha}{\alpha^{2}}\int_{\beta_{min}}^{1-\alpha}d\beta\beta(1+\alpha+\beta)\bigg],\nonumber
\end{eqnarray}
with $r(m_{Q},s)$ defined as $(\alpha+\beta)m_{Q}^2-\alpha\beta s$. The
integration limits are given by
$\alpha_{min}=\Big(1-\sqrt{1-4m_{Q}^{2}/s}\Big)/2$,
$\alpha_{max}=\Big(1+\sqrt{1-4m_{Q}^{2}/s}\Big)/2$, and
$\beta_{min}=\alpha m_{Q}^{2}/(s\alpha-m_{Q}^{2})$.
The term $\hat{B}\Pi_{1}^{\mbox{cond}}$ reads
\begin{eqnarray}
\hat{B}\Pi_{1}^{\mbox{cond}}&=&\frac{\langle\bar{q}q\rangle\langle
g^{2}G^{2}\rangle}{3\cdot2^{9}\pi^{4}}m_{Q}^{3}\int_{0}^{1}d\alpha\bigg[\frac{1}{\alpha^{3}}\int_{0}^{1-\alpha}d\beta(\alpha+\beta)(1+\alpha+\beta)e^{-\frac{(\alpha+\beta)m_{Q}^{2}}{\alpha\beta M^{2}}}-\frac{1}{1-\alpha}e^{-\frac{m_{Q}^{2}}{\alpha(1-\alpha)M^{2}}}\bigg]
\nonumber\\&&-\frac{\langle\bar{q}q\rangle\langle g\bar{q}\sigma\cdot G q\rangle}{2^{5}\pi^{2}}m_{Q}^{2}\int_{0}^{1}d\alpha\int_{0}^{1-\alpha}d\beta\bigg[1+\frac{(\alpha+\beta)m_{Q}^{2}}{\alpha\beta M^{2}}\bigg]e^{-\frac{(\alpha+\beta)m_{Q}^{2}}{\alpha\beta M^{2}}}\nonumber\\&&
+\frac{\langle g^{2}G^{2}\rangle^{2}}{3^{2}\cdot2^{15}\pi^{6}}m_{Q}^{4}\int_{0}^{1}\frac{d\alpha}{\alpha^{2}}\int_{0}^{1-\alpha}\frac{d\beta}{\beta^{2}}(1-\alpha-\beta)(1+\alpha+\beta)\frac{1}{M^{2}}e^{-\frac{(\alpha+\beta)m_{Q}^{2}}{\alpha\beta M^{2}}}\nonumber\\&&
+\frac{\langle\bar{q}q\rangle\langle
g^{3}G^{3}\rangle}{3\cdot2^{11}\pi^{4}}m_{Q}\int_{0}^{1}\frac{d\alpha}{\alpha^{4}}\int_{0}^{1-\alpha}d\beta(1+\alpha+\beta)\bigg[\alpha(\alpha+6\beta)-\frac{2(\alpha+\beta)m_{Q}^{2}}{M^{2}}\bigg]e^{-\frac{(\alpha+\beta)m_{Q}^{2}}{\alpha\beta M^{2}}}\nonumber\\&&
+\frac{\langle g^{2}G^{2}\rangle\langle g\bar{q}\sigma\cdot G q\rangle}{3\cdot2^{11}\pi^{4}}m_{Q}\int_{0}^{1}\frac{d\alpha}{\alpha^{3}}\bigg\{2\Big[3\alpha(1-\alpha)-\frac{m_{Q}^{2}}{M^{2}}\Big]e^{-\frac{m_{Q}^{2}}{\alpha(1-\alpha)M^{2}}}\nonumber\\&&
+\int_{0}^{1-\alpha}d\beta\Big[-3\alpha\beta+(\alpha+\beta)\frac{m_{Q}^{2}}{M^{2}}\Big]e^{-\frac{(\alpha+\beta)m_{Q}^{2}}{\alpha\beta M^{2}}}\bigg\},\nonumber
\end{eqnarray}
with $\langle\bar{q}q\rangle\langle g\bar{q}\sigma\cdot G q\rangle$, $\langle g^{2}G^{2}\rangle^{2}$, $\langle\bar{q}q\rangle\langle
g^{3}G^{3}\rangle$, and $\langle g^{2}G^{2}\rangle\langle g\bar{q}\sigma\cdot G q\rangle$ denoting the
two-quark multiply mixed condensate, four-gluon condensate, two-quark multiply three-gluon
condensate, and two-gluon multiply mixed condensate, respectively.
%%%%%%%%%%%%%%%%%%%%%%%%%%%%%%%%%%%%%%%%%%%%%%%%%%%%%%%%%%%%%%%%%%%
\section{Numerical analysis and discussions}\label{sec3}
In this section, the sum rule (\ref{sum rule
1}) will be numerically analyzed. The input values are taken as $m_{c}=1.23\pm0.05~\mbox{GeV}$,
$\langle\bar{q}q\rangle=-(0.23\pm0.03)^{3}~\mbox{GeV}^{3}$, $\langle
g\bar{q}\sigma\cdot G q\rangle=m_{0}^{2}~\langle\bar{q}q\rangle$,
$m_{0}^{2}=0.8\pm0.1~\mbox{GeV}^{2}$, $\langle
g^{2}G^{2}\rangle=0.88~\mbox{GeV}^{4}$, and $\langle
g^{3}G^{3}\rangle=0.045~\mbox{GeV}^{6}$ \cite{overview2}.
In the standard
procedure of sum rule analysis,
one should analyze the OPE convergence
and the pole contribution dominance to determine the conventional Borel window for $M^2$:
 on the one side, the lower
constraint for $M^{2}$ is obtained by considering that the
perturbative contribution should be larger than each condensate
contribution to have a good
convergence in the
OPE side; on the other
side, the upper bound for $M^{2}$ is obtained by the consideration
that the pole contribution should be larger than the continuum
state contributions. At the same time, the threshold
$\sqrt{s_{0}}$ is not arbitrary but characterizes the
beginning of continuum states. Hence,
the most expected case is that
one could naturally find the conventional Borel windows for studied states
to make QCD sum rules work well.

In order to test the
convergence of OPE, its various contributions, i.e. the perturbative,
two-quark, two-gluon, mixed, four-quark,  three-gluon, two-quark
multiply two-gluon,  two-quark multiply mixed, four-gluon, two-quark multiply three-gluon,
and two-gluon multiply
mixed condensate contributions, are compared as a function of $M^2$ and showed
in FIG. 1.
Graphically, one could see that in the OPE side there exists some similar problem
which has been discussed in some of our recent works \cite{Zs} and
others e.g. \cite{Matheus,HXChen,ZGWang}.
Concretely, here some condensates especially two-quark condensate $\langle\bar{q}q\rangle$
are very large and play an important role in the OPE side,
which makes the standard OPE convergence (i.e. the perturbative at least
larger than each condensate contribution) happen only at very large values of $M^2$.
The consequence is that it is difficult to find a conventional Borel window
where both the pole dominates over the continuum and the OPE converges well.
Following the similar treatment in Refs. \cite{Zs},
we could try
releasing the rigid convergence criterion of the perturbative contribution
larger than each condensate contribution here.
It is not too bad for the present case, there are two main condensates
i.e. $\langle\bar{q}q\rangle$ and $\langle
g\bar{q}\sigma\cdot G q\rangle$
and they could cancel out each other to some extent
since they have different signs. What is also very important,
most of other condensates calculated
are very small,
which means that they could not radically influence the character of OPE convergence.
Therefore, one could find that
the OPE convergence is still under control here.
In the phenomenological side,
the comparison between pole and
continuum contributions of sum rule (\ref{sumrule1}) as a function of the
Borel parameter $M^2$ for the threshold value
$\sqrt{s_{0}}=4.4~\mbox{GeV}$ is shown in FIG. 2, which shows that the relative pole
contribution is approximate to $50\%$ at $M^{2}=2.7~\mbox{GeV}^{2}$
and decreases with $M^{2}$. Similarly, the upper bound values of Borel parameters are
$M^{2}=2.6~\mbox{GeV}^{2}$ for
$\sqrt{s_0}=4.3~\mbox{GeV}$ and $M^{2}=2.9~\mbox{GeV}^{2}$ for
$\sqrt{s_0}=4.5~\mbox{GeV}$.
Thus, the Borel window for
$\bar{D}D^{*}$ is taken as $M^{2}=2.1\sim2.7~\mbox{GeV}^{2}$ for
$\sqrt{s_0}=4.4~\mbox{GeV}$. Similarly, the proper range of $M^{2}$
are $2.1\sim2.6~\mbox{GeV}^{2}$ for $\sqrt{s_0}=4.3~\mbox{GeV}$ and
$2.1\sim2.9~\mbox{GeV}^{2}$ for $\sqrt{s_0}=4.5~\mbox{GeV}$. The mass of the $\bar{D}D^{*}$ molecular state as
a function of $M^2$ from sum rule (\ref{sum rule 1}) is
shown in FIG. 3 and it is numerically calculated to be $3.86\pm0.13~\mbox{GeV}$ in the above chosen work windows.
Considering the uncertainty rooting in the variation of quark masses and
condensates, we gain
$3.86\pm0.13\pm0.14~\mbox{GeV}$ (the
first error reflects the uncertainty due to variation of $\sqrt{s_{0}}$
and $M^{2}$, and the second error resulted from the variation of
QCD parameters) or concisely $3.86\pm0.27~\mbox{GeV}$
for the $S$-wave $\bar{D}D^{*}$.

\begin{figure}[htb!]
\centerline{\epsfysize=6.8truecm\epsfbox{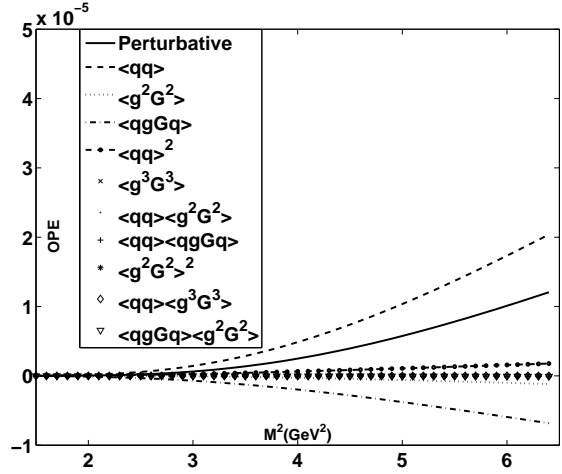}}
\caption{The OPE contribution in sum rule
(\ref{sumrule1}) for $\sqrt{s_{0}}=4.4~\mbox{GeV}$.
The OPE convergence is shown by comparing the perturbative, two-quark condensate, two-gluon condensate, mixed condensate, four-quark condensate,  three-gluon
condensate, two-quark
multiply two-gluon condensate,  two-quark multiply mixed
condensate, four-gluon condensate, two-quark multiply three-gluon condensate,
mixed multiply
two-gluon condensate contributions.}
\end{figure}

\begin{figure}
\centerline{\epsfysize=6.8truecm\epsfbox{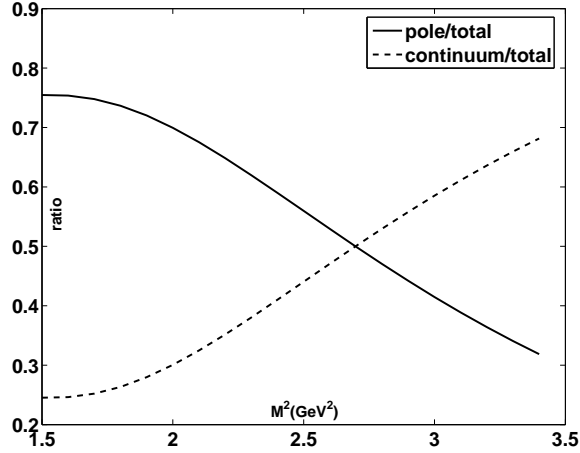}}
\caption{The phenomenological contribution in sum rule
(\ref{sumrule1}) for $\sqrt{s_{0}}=4.4~\mbox{GeV}$.
The solid line is the relative pole contribution (the pole
contribution divided by the total, pole plus continuum contribution)
as a function of $M^2$ and the dashed line is the relative continuum
contribution.}
\end{figure}

\begin{figure}
\centerline{\epsfysize=6.8truecm
\epsfbox{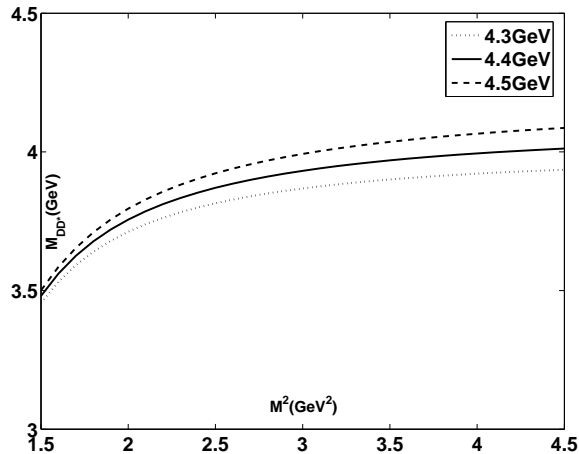}}\caption{
The mass of the $\bar{D}D^{*}$ molecular state as
a function of $M^2$ from sum rule (\ref{sum rule 1}). The continuum
thresholds are taken as $\sqrt{s_0}=4.3\sim4.5~\mbox{GeV}$. The
ranges of $M^{2}$ is $2.1\sim2.6~\mbox{GeV}^{2}$ for
$\sqrt{s_0}=4.3~\mbox{GeV}$, $2.1\sim2.7~\mbox{GeV}^{2}$ for
$\sqrt{s_0}=4.4~\mbox{GeV}$, and $2.1\sim2.9~\mbox{GeV}^{2}$ for
$\sqrt{s_0}=4.5~\mbox{GeV}$.}
\end{figure}

\section{Summary}\label{sec4}
Stimulated by the newly observed charged charmonium-like
structure $Z_{c}(3900)$ for which can not be simple $c\bar{c}$ conventional mesons
and are some ideal candidates for exotic hadrons, we present an improved QCD sum rule study of our previous work
on the $\bar{D}D^{*}$ molecular state to
investigate that whether it could be a $S$-wave $\bar{D}D^{*}$ molecular state.
In order to ensure the quality of QCD sum rule analysis, contributions of
up to dimension nine are calculated to test the convergence of OPE.
We find that some condensates in particular $\langle\bar{q}q\rangle$
play an important role and make the standard OPE convergence (i.e. the perturbative at least
larger than each condensate contribution) happen at very large values of Borel parameters $M^{2}$.
By releasing the rigid OPE convergence criterion, one could find that the OPE convergence is still under control
and the final result $3.86\pm0.27~\mbox{GeV}$ is obtained for the $S$-wave $\bar{D}D^{*}$ molecular state,
which coincides with the experimental data of $Z_{c}(3900)$.
From the final result, one could assuredly state
that it could provide some support to
the $\bar{D}D^{*}$ molecular explanation of $Z_{c}(3900)$.
At the same time, one should note that the $\bar{D}D^{*}$ molecular state is just one
possible theoretical interpretation of $Z_{c}(3900)$, and
it does not mean that one could arbitrarily excluded
some other possible explanations (e.g. tetraquark states) at the present time
just from the result here.
In fact, more minute information on the nature structures of $Z_{c}(3900)$
could be revealed by the future contributions of both experimental observations
and theoretical studies.

\emph{Note added}--As we were preparing to submit this
paper, we became aware of a paper from our colleague that also
analyzes $Z_{c}(3900)$ as a $\bar{D}D^{*}$ molecular state
with QCD sum rules \cite{Cui}, but then they
consider contributions up to the same dimension six
as our previous work \cite{Previous}.

%%%%%%%%%%%%%%%%%%%%%%%%%%%%%%%%%%%%%%
\begin{acknowledgments}
The author thanks Xiang Liu, BeiJiang Liu, and Qiang Zhao for the interesting
discussions on $Z_{c}(3900)$ during the second international conference on QCD and Hadron Physics
held at the IMP of the Chinese Academy of Science.
The author would also like to acknowledge PengMing Zhang for his
effective organization in that conference,
in which part of the work was done.
This work was supported by the National Natural Science
Foundation of China under Contract Nos. 11105223, 10947016, 10975184, and the
Foundation of NUDT (No. JC11-02-12).
\end{acknowledgments}
%%%%%%%%%%%%%%%%%%%%%%%%%%%%%%%%%%%%%%%%%%%%%%%%%%%%

\end{document}